\documentclass[twocolumn,showpacs,aps,prl,superscriptaddress]{revtex4}

\newcommand{\SLACPubNumber} {13133}

\long\def\inst#1{\par\nobreak\kern 4pt\nobreak
    {\it #1}\par\vskip 10pt plus 3pt minus 3pt}

\usepackage[dvips]{graphicx}
\usepackage{epsf}  
\usepackage{epsfig}  
\usepackage{rotating}
\usepackage{color}
\usepackage{colordvi}
\usepackage{mdwlist}

\newcommand{\psfile}[3][]{ 
  \begin{center}
    \setlength{\epsfxsize}{#3\linewidth}\leavevmode
    \def\noOpt{}\def\testit{#1}\ifx\testit\noOpt%
      \epsfbox{#2}%
    \else%
      \epsfbox[#1]{#2}%
    \fi
  \end{center}
}

\RequirePackage{xspace}

\usepackage{relsize}
\def\babar{\mbox{\slshape B\kern-0.1em{\smaller A}\kern-0.1em
    B\kern-0.1em{\smaller A\kern-0.2em R}}}

\def\epem       {\ensuremath{e^+e^-}\xspace}

\def\mumu       {\ensuremath{\mu^+\mu^-}\xspace}

\def\ellell     {\ensuremath{\ell^+ \ell^-}\xspace}

\def\piz   {\ensuremath{\pi^0}\xspace}

\def\pip   {\ensuremath{\pi^+}\xspace}
\def\pim   {\ensuremath{\pi^-}\xspace}

\def\Kbar  {\kern 0.2em\overline{\kern -0.2em K}{}\xspace}

\def\Kz    {\ensuremath{K^0}\xspace}
\def\Kzb   {\ensuremath{\Kbar^0}\xspace}
\def\KzKzb {\ensuremath{\Kz \kern -0.16em \Kzb}\xspace}
\def\Kp    {\ensuremath{K^+}\xspace}
\def\Km    {\ensuremath{K^-}\xspace}

\def\KpKm  {\ensuremath{\Kp \kern -0.16em \Km}\xspace}
\def\KS    {\ensuremath{K^0_{\scriptscriptstyle S}}\xspace}

\def\Kstar   {\ensuremath{K^*}\xspace}

\def\Dbar    {\kern 0.2em\overline{\kern -0.2em D}{}\xspace}

\def\Dz      {\ensuremath{D^0}\xspace}
\def\Dzb     {\ensuremath{\Dbar^0}\xspace}
\def\DzDzb   {\ensuremath{\Dz {\kern -0.16em \Dzb}}\xspace}
\def\Dp      {\ensuremath{D^+}\xspace}
\def\Dm      {\ensuremath{D^-}\xspace}

\def\DpDm    {\ensuremath{\Dp {\kern -0.16em \Dm}}\xspace}

\def\B       {\ensuremath{B}\xspace}
\def\Bbar    {\kern 0.18em\overline{\kern -0.18em B}{}\xspace}

\def\BB      {\ensuremath{B\Bbar}\xspace}
\def\Bz      {\ensuremath{B^0}\xspace}
\def\Bzb     {\ensuremath{\Bbar^0}\xspace}
\def\BzBzb   {\ensuremath{\Bz {\kern -0.16em \Bzb}}\xspace}
\def\Bu      {\ensuremath{B^+}\xspace}
\def\Bub     {\ensuremath{B^-}\xspace}

\def\BpBm    {\ensuremath{\Bu {\kern -0.16em \Bub}}\xspace}

\def\BorBbar    {\kern 0.18em\optbar{\kern -0.18em B}{}\xspace}
\def\DorDbar    {\kern 0.18em\optbar{\kern -0.18em D}{}\xspace}
\def\KorKbar    {\kern 0.18em\optbar{\kern -0.18em K}{}\xspace}

\mathchardef\Upsilon="7107
\def\Y#1S{\ensuremath{\Upsilon{(#1S)}}\xspace}

\def\FourS {\Y4S}

\mathchardef\Deltares="7101
\mathchardef\Xi="7104
\mathchardef\Lambda="7103
\mathchardef\Sigma="7106
\mathchardef\Omega="710A

\def\Deltabar{\kern 0.25em\overline{\kern -0.25em \Deltares}{}\xspace}
\def\Lbar{\kern 0.2em\overline{\kern -0.2em\Lambda\kern 0.05em}\kern-0.05em{}\xspace}
\def\Sigbar{\kern 0.2em\overline{\kern -0.2em \Sigma}{}\xspace}
\def\Xibar{\kern 0.2em\overline{\kern -0.2em \Xi}{}\xspace}
\def\Obar{\kern 0.2em\overline{\kern -0.2em \Omega}{}\xspace}
\def\Nbar{\kern 0.2em\overline{\kern -0.2em N}{}\xspace}
\def\Xb{\kern 0.2em\overline{\kern -0.2em X}{}\xspace}

\def\mes        {\mbox{$m_{\rm ES}$}\xspace}

\def\DeltaE     {\mbox{$\Delta E$}\xspace}

\newcommand{\tev}{\ensuremath{\mathrm{\,Te\kern -0.1em V}}\xspace}
\newcommand{\gev}{\ensuremath{\mathrm{\,Ge\kern -0.1em V}}\xspace}
\newcommand{\mev}{\ensuremath{\mathrm{\,Me\kern -0.1em V}}\xspace}
\newcommand{\kev}{\ensuremath{\mathrm{\,ke\kern -0.1em V}}\xspace}
\newcommand{\ev}{\ensuremath{\mathrm{\,e\kern -0.1em V}}\xspace}
\newcommand{\gevc}{\ensuremath{{\mathrm{\,Ge\kern -0.1em V\!/}c}}\xspace}
\newcommand{\mevc}{\ensuremath{{\mathrm{\,Me\kern -0.1em V\!/}c}}\xspace}
\newcommand{\gevcc}{\ensuremath{{\mathrm{\,Ge\kern -0.1em V\!/}c^2}}\xspace}
\newcommand{\mevcc}{\ensuremath{{\mathrm{\,Me\kern -0.1em V\!/}c^2}}\xspace}

\def\mus  {\ensuremath{\rm \,\mus}\xspace}

\def\mus        {\ensuremath{\,\mu{\rm s}}\xspace}    

\def\to                 {\ensuremath{\rightarrow}\xspace}

\def\pep2{PEP-II}

\def\gsim{{~\raise.15em\hbox{$>$}\kern-.85em
          \lower.35em\hbox{$\sim$}~}\xspace}
\def\lsim{{~\raise.15em\hbox{$<$}\kern-.85em
          \lower.35em\hbox{$\sim$}~}\xspace}

\xspace

\newcommand{\epjBase}        {Eur.\ Phys.\ Jour.\xspace}

\newcommand{\jprBase}        {Phys.\ Rev.\xspace}
\newcommand{\jplBase}        {Phys.\ Lett.\xspace}
\newcommand{\nimBaseA}       {Nucl.\ Instrum.\ Methods Phys.\ Res., Sect.\ A\xspace}

\newcommand{\nimBaseC}       {Nucl.\ Instrum.\ Methods Phys.\ Res., Sect.\ C\xspace}

\newcommand{\npBase}         {Nucl.\ Phys.\xspace}
\newcommand{\zpBase}         {Z.\ Phys.\xspace}

\newcommand{\epjc}      [1]  {\epjBase\ C~{\bf #1}}

\newcommand{\mpl}       [1]  {{Mod.\ Phys.\ Lett.\ {\bf #1}}}

\newcommand{\nim}       [1]  {\nimBaseC~{\bf #1}}

\newcommand{\nima}      [1]  {\nimBaseA~{\bf #1}}

\newcommand{\npb}       [1]  {\npBase\ B~{\bf #1}}

\newcommand{\npbps}     [1]  {{Nucl.\ Phys.\ B~Proc.\ Suppl.\ {\bf #1}}}

\newcommand{\plb}       [1]  {\jplBase\ B~{\bf #1}}

\newcommand{\pr}        [1]  {\jprBase\ {\bf #1}}

\newcommand{\progtp}    [1]  {{Prog.\ Theor.\ Phys.\ {\bf #1}}}

\newcommand{\zpc}       [1]  {\zpBase\ C~{\bf #1}}

\def\jetset74   {\mbox{\tt Jetset \hspace{-0.5em}7.\hspace{-0.2em}4}\xspace}

\newcommand{\gevcccc}{\ensuremath{{\mathrm{\,Ge\kern -0.1em V^2\!/}c^4}}\xspace}

\def\ththr {\ensuremath{\theta_{\rm thrust}}\xspace}

\def\emu {\ensuremath{e\mu}\xspace}

\def\ctl {\ensuremath{\cos\theta_{\ell}}\xspace}

\def\ctlsq {\ensuremath{\cos^2\theta_{\ell}}\xspace}

\def\ctk {\ensuremath{\cos\theta_{K}}\xspace}

\def\ththrroe {\mbox{$\theta_{\rm thrust}^{\rm ROE}$}\xspace}
\def\afb {\mbox{${\cal A}_{FB}$}\xspace}
\def\fl {\mbox{$F_L$}\xspace}

\def\modeeightshort {\ensuremath{\Kp \pim \mumu}\xspace}
\def\modeelevenshort {\ensuremath{\KS \pip \epem}\xspace}

\def\modesevenshort {\ensuremath{\KS \pip \mumu}\xspace}
\def\modesixshort {\ensuremath{\Kp \piz \mumu}\xspace}
\def\modetenshort {\ensuremath{\Kp \piz \epem}\xspace}

\def\modetwelveshort {\ensuremath{\Kp \pim \epem}\xspace}

\def\mue       {\ensuremath{\mu^+e^-}\xspace}
\def\emu       {\ensuremath{e^+\mu^-}\xspace}

\usepackage{relsize}
\usepackage{xspace}

\begin{document}

{\pagestyle{empty}

\begin{flushleft}
\babar-PUB-08/022 \\
SLAC-PUB-\SLACPubNumber \\
arXiv:0804.4412 [hep-ex]
\end{flushleft}

\title{
        {\mathversion{bold}
         Angular Distributions in the Decays $\B \rightarrow \Kstar\ellell$}
}

%
\author{B.~Aubert}
\author{M.~Bona}
\author{Y.~Karyotakis}
\author{J.~P.~Lees}
\author{V.~Poireau}
\author{X.~Prudent}
\author{V.~Tisserand}
\author{A.~Zghiche}
\affiliation{Laboratoire de Physique des Particules, IN2P3/CNRS et Universit\'e de Savoie, F-74941 Annecy-Le-Vieux, France }
\author{J.~Garra~Tico}
\author{E.~Grauges}
\affiliation{Universitat de Barcelona, Facultat de Fisica, Departament ECM, E-08028 Barcelona, Spain }
% ...
\collaboration{The \babar\ Collaboration}
\noaffiliation

\begin{abstract}
We use a sample of 384 million $\BB$ events collected with
the \babar\, detector at the \pep2\ $\epem$ collider to study
angular distributions in the rare decays $\B \rightarrow \Kstar\ellell$,
where $\ell^+\ell^-$ is either $e^+e^-$ or $\mu^+\mu^-$.
For low dilepton invariant masses, $m_{\ell\ell}<2.5\gevcc$,
we measure a lepton forward-backward asymmetry
$\afb=0.24^{+0.18}_{-0.23} \pm 0.05$
and $K^*$ longitudinal polarization
$\fl=0.35 \pm 0.16 \pm 0.04$.
For $m_{\ell\ell}>3.2\gevcc$,
we measure $\afb=0.76^{+0.52}_{-0.32} \pm 0.07$
and $\fl=0.71^{+0.20}_{-0.22} \pm 0.04$.
\end{abstract}


\vfill

\maketitle
}

\pagestyle{plain}

\begin{figure}[b!]
\begin{center}
\includegraphics[height=3.5cm]{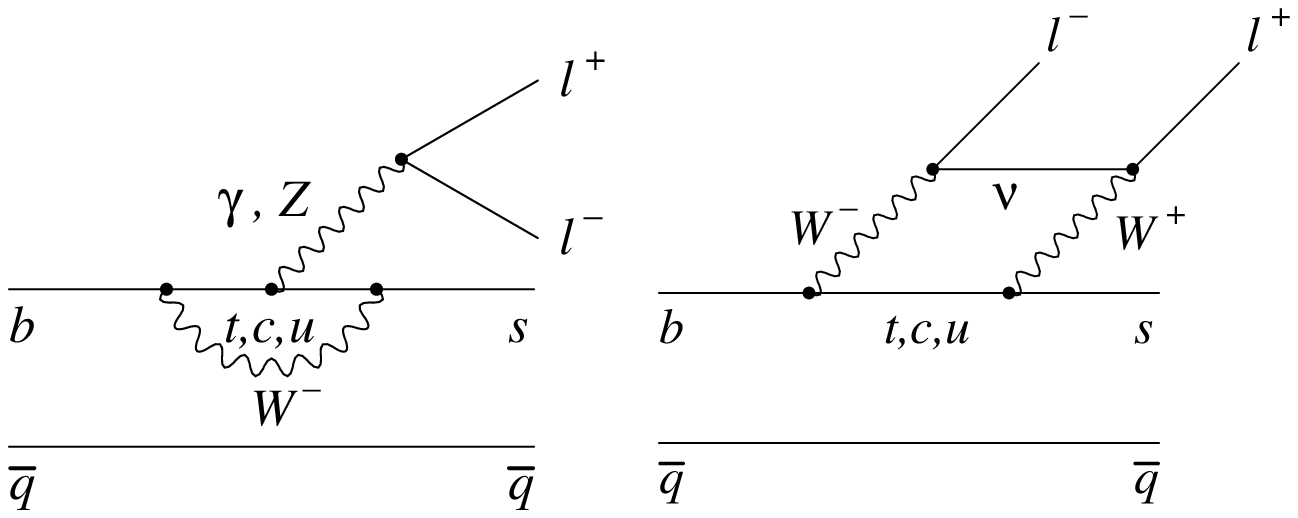}
\caption{Lowest-order Feynman diagrams for $b \to s \ellell$.}
\label{fig:sll_diagrams}
\end{center}
\end{figure}

The decays $B \rightarrow K^* \ellell$,
where $K^*\to K\pi$ and $\ellell$ is
either an $\epem$ or $\mumu$ pair,
arise from flavor-changing neutral
currents (FCNC), which are forbidden
at tree level in the Standard Model (SM).
The lowest-order SM processes contributing to these decays
are the photon or $Z$ penguin and the $W^+W^-$ box diagrams
shown in Fig.~\ref{fig:sll_diagrams}. The amplitudes can be expressed
in terms of effective Wilson coefficients for the electromagnetic penguin,
$C_{7}^{\textrm{\tiny{eff}}}$, and the vector and axial-vector electroweak contributions,
$C_{9}^{\textrm{\tiny{eff}}}$ and $C_{10}^{\textrm{\tiny{eff}}}$ respectively,
arising from the interference of the $Z$ penguin
and $W^+W^-$ box diagrams~\cite{Buchalla}.
The angular distributions
in these decays as a function of dilepton
mass squared $q^2=m_{\ellell}^2$ are sensitive to
many possible new physics contributions
\cite{NewPhysics}.

We describe measurements of the distribution of the angle
$\theta_K$ between the $K$ and the $B$ directions in the $K^*$ rest frame.
A fit to $\ctk$ of the form~\cite{KrugerMatias}
\begin{equation}
\frac{3}{2} \fl \cos^2\theta_K + \frac{3}{4}(1-\fl)(1-\cos^2\theta_K)
\end{equation}
determines $\fl$, the $K^*$ longitudinal polarization fraction.
We also describe measurements of the distribution of the angle
$\theta_{\ell}$ between the $\ell^+(\ell^-)$ and the $B(\Bbar)$ direction
in the $\ell^+\ell^-$ rest frame.
A fit to $\ctl$ of the form~\cite{KrugerMatias}
\begin{equation}
{{3}\over{4}}\fl (1-\ctlsq) + {{3}\over{8}}(1-\fl )(1+\ctlsq) + \afb \ctl
\end{equation}
determines $\afb$, the lepton forward-backward asymmetry.
These measurements
are done in a low $q^2$ region $0.1 < q^2 < 6.25\gevcccc$, and in a high
$q^2$ region above $10.24\gevcccc$. We remove the $J/\psi$ and $\psi(2S)$ resonances
by vetoing events in the regions $q^2=6.25$-$10.24\gevcccc$ and $q^2=12.96$-$14.06\gevcccc$
respectively.

\begin{figure}[tb!]
\begin{center}
\includegraphics[width=1.0\linewidth]{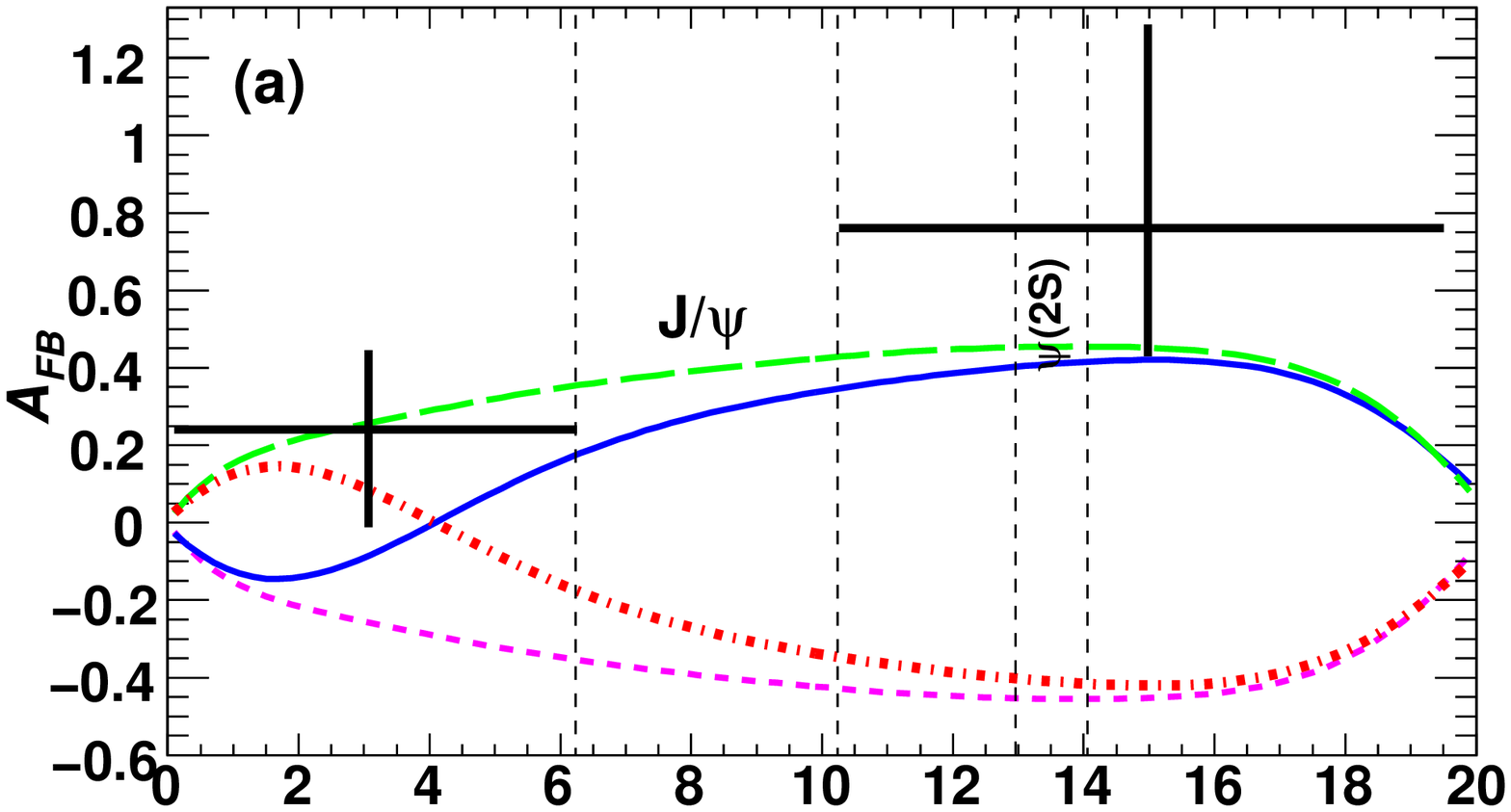}
\includegraphics[width=1.0\linewidth]{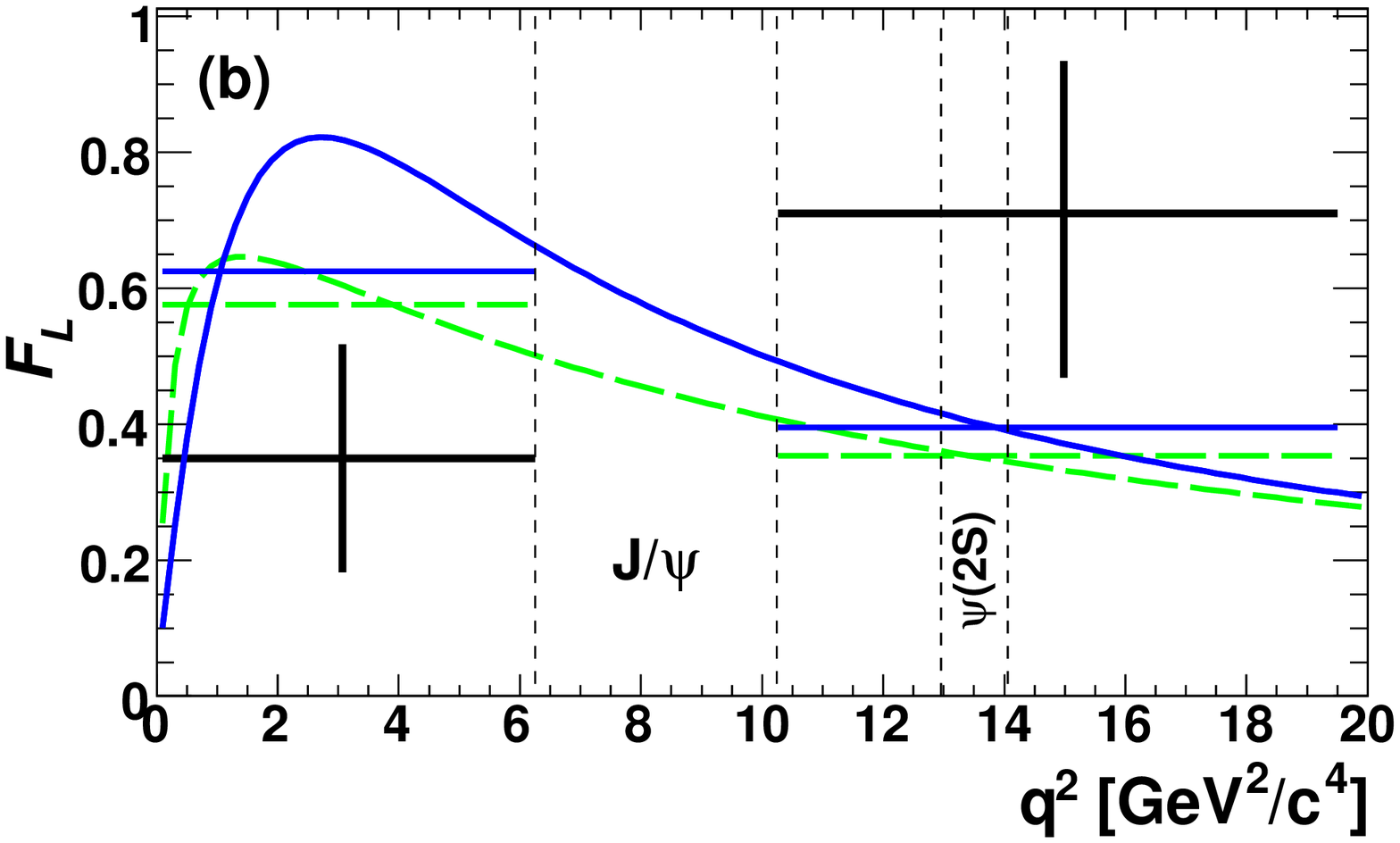}
\caption{Plots of our results for (a) $\afb$ and (b) $F_{L}$
for the decay $B\to K^* \ell^+\ell^-$ showing comparisons
with SM (solid); $C_7^{\textrm{\tiny{eff}}} = -C_7^{\textrm{\tiny{eff}}}$ (long dash);
$C_9^{\textrm{\tiny{eff}}} C_{10}^{\textrm{\tiny{eff}}} = -C_9^{\textrm{\tiny{eff}}} C_{10}^{\textrm{\tiny{eff}}}$ (short dash);
$C_7^{\textrm{\tiny{eff}}} = -C_7^{\textrm{\tiny{eff}}}, C_9^{\textrm{\tiny{eff}}} C_{10}^{\textrm{\tiny{eff}}} = -C_9^{\textrm{\tiny{eff}}} C_{10}^{\textrm{\tiny{eff}}}$ (dash-dot).
Statistical and systematic errors are added in quadrature.
Expected $F_{L}$ values integrated over each $q^2$ region are also shown.
The $F_{L}$ curves with $C_9^{\textrm{\tiny{eff}}} C^{\textrm{\tiny{eff}}} = -C_9^{\textrm{\tiny{eff}}} C_{10}^{\textrm{\tiny{eff}}}$
are nearly identical to the two curves shown.}
\label{fig:newafbfl}
\end{center}
\end{figure}

The SM predicts a distinctive variation
of $\afb$ arising from the interference
between the different amplitudes.
The expected SM dependence of $A_{FB}$
and $F_{L}$ on $q^2$ along with variations
due to opposite-sign Wilson coefficients
are shown in Fig.~\ref{fig:newafbfl}.
At low $q^2$, where $C_7^{\textrm{\tiny{eff}}}$ dominates, $\afb$ is expected to be small
with a zero-crossing point at $q^2 \sim 4 \gevcccc$~\cite{Ali01,AFB_SM,bsgamma_beneke}.
There is an experimental constraint on the magnitude of $C_7^{\textrm{\tiny{eff}}}$ coming
from the branching fraction for $b\to s\gamma$~\cite{bsgamma_beneke,bsgamma}, which corresponds to
the limit $q^2 \to 0$. However, a reversal of the sign of $C_7^{\textrm{\tiny{eff}}}$ is allowed.
At high $q^2$, the product of $C_9^{\textrm{\tiny{eff}}}$ and $C_{10}^{\textrm{\tiny{eff}}}$
is expected to give a large positive asymmetry.
Right-handed weak currents have an opposite-sign $C_9^{\textrm{\tiny{eff}}}C_{10}^{\textrm{\tiny{eff}}}$ which
would give a negative $\afb$ at high $q^2$.
Contributions from non-SM processes can change the magnitudes and relative signs
of $C_7^{\textrm{\tiny{eff}}}$, $C_9^{\textrm{\tiny{eff}}}$ and $C_{10}^{\textrm{\tiny{eff}}}$, and may introduce complex phases between
them~\cite{KrugerMatias,Hovhannisyan}.
An experimental determination of $\fl$ is required to obtain a
model-independent $\afb$ result, and thus avoid drawing possibly
incorrect inferences about new physics from our observations.

We reconstruct signal events in six separate
flavor-specific final states
containing an $\epem$ or $\mumu$ pair, and a $\Kstar(892)$
candidate reconstructed as $K^+ \pim$, $K^+ \piz$ or
$\KS \pip$ (or their charge conjugates).
To understand combinatorial backgrounds we also reconstruct
samples containing the same hadronic final states and
$e^{\pm}\mu^{\mp}$ pairs, where no signal is expected
because of lepton flavor conservation.
To understand backgrounds from hadrons ($h$)
misidentified as muons, we similarly reconstruct
samples containing $h^{\pm}\mu^{\mp}$
pairs with no particle identification
requirement for the $h^{\pm}$.

We use a dataset of $384$ million $\BB$ pairs
collected at the $\FourS$ resonance
with the \babar\ detector \cite{BaBarDetector} at
the \pep2\ asymmetric-energy $\epem$ collider.
Tracking is provided
by a five-layer silicon vertex tracker and a 40-layer drift chamber
in a 1.5~T magnetic field.
We identify electrons with a CsI(Tl) electromagnetic calorimeter,
muons with an instrumented magnetic flux return, and
$K^+$ using a detector of internally reflected
Cherenkov light as well as ionization energy loss information.
Charged tracks other than identified $e$, $\mu$ and $K$ candidates are treated as pions.
Electrons (muons) are required to have momenta $p > 0.3 (0.7) \gevc$
in the laboratory frame.
We add photons to electrons when they are consistent with bremsstrahlung,
and do not use electrons that arise from photon conversions to low-mass $\epem$ pairs.
Neutral $\KS \to \pip \pim$ candidates are required to have an invariant mass
consistent with the nominal $K^0$ mass~\cite{JPsi_BF}, and a flight distance
from the $\epem$ interaction point which is more than three times
its uncertainty. Neutral pion candidates are formed from
two photons with $E_{\gamma} > 50 \mev$, and an invariant mass
between $115$ and $155 \mevcc$.
We require $\Kstar(892)$ candidates to have an invariant
mass $0.82 < M(K\pi) < 0.97 \gevcc$.

$B\to K^*\ell^+\ell^-$ decays are characterized by the kinematic
variables $\mes=\sqrt{s/4 -p^{*2}_B}$ and
$\Delta E = E_B^* - \sqrt{s}/2$, where $p^*_B$ and $E_B^*$ are
the reconstructed $B$ momentum and energy in the center-of-mass (CM) frame,
and $\sqrt{s}$ is the total CM energy.
We define a fit region $\mes > 5.2 \gevcc$, with
$-0.07<\Delta E<0.04$ ($-0.04<\Delta E<0.04$) $\gev$ for
$e^+e^-$ ($\mu^+\mu^-$) final states in the low $q^2$ region, and
$-0.08<\Delta E<0.05$ ($-0.05<\Delta E<0.05$) $\gev$ for high $q^2$.
We use the wider (narrower) $\Delta E$ windows to select the $e^{\pm}\mu^{\mp}$
($h^{\pm}\mu^{\mp}$) background samples.

The most significant background arises from random combinations of
leptons from semileptonic $B$ and $D$ decays.
In $B\Bbar$ events the leptons are kinematically correlated
if they come from $B\to D^{(*)}\ell\nu$, $D\to K^{(*)}\ell\nu$.
Uncorrelated backgrounds combine leptons from separate $B$ decays
or from continuum $\epem \to c\bar{c}$ events.
We suppress these types of combinatorial background
through the use of neural networks (NN).
For each final state we use four separate NN
designed to suppress either continuum or $B\Bbar$ backgrounds
in either the low or high $q^2$ regions, and
different selections of NN inputs are used
depending on $q^2$ bin (low, high), the identity of the
leptons in the final state ($e$, $\mu$), and the type
of background ($B\Bbar$, continuum). Inputs include:
\begin{itemize*}
\item event thrust;
\item ratio of second-to-zeroth Fox-Wolfram moments~\cite{Fox:1978vw};
\item $\mes$ and $\DeltaE$ of the rest of the event (ROE),
  comprising all charged tracks and neutral energy deposits
  not used to reconstruct the signal candidate;
\item the magnitude of the total event transverse momentum, which is correlated
with missing energy due to unreconstructed neutrinos in background
semileptonic decays;
\item di-lepton system's distance of closest approach along
  the z-axis, and separately in the xy-plane, to the primary interaction point;
\item vertex probability of the signal candidate and,
  separately, of the di-lepton system;
\item the cosines in the CM frame of
  the angle between the \B candidate's momentum and the $z$ axis,
  the angle between the event thrust axis and the $z$ axis (\ththr),
  the angle between the ROE thrust axis and the $z$ axis (\ththrroe),
  and the angle between \ththrroe and \ththr.
\end{itemize*}

There is also a background contribution in the signal region from $B \to D(\Kstar \pi) \pi$ decays,
where both pions are misidentified.
The misidentification rates for muons and electrons are $\sim 2\%$ and $\sim 0.1\%$, respectively,
so this background is only significant in the $\mu^+\mu^-$ final states.
These events are vetoed if the invariant mass of the $\Kstar\pi$ system is
in the range $1.84$-$1.90 \gevcc$.

We optimize the NN and $\Delta E$ selections for each final state
in each $q^2$ bin to give the best combined statistical signal significance
in the $\mes$ signal region $\mes>5.27 \gevcc$ for the sum of all
six final states.
After all these selections have been applied, the final reconstruction
efficiencies and expected yields for signal events (calculated using world
average branching fractions~\cite{bsgamma}), as well as expected yields for background
events in the signal region, are shown in Table~\ref{tab:effs}.

\begin{table}[tb!]
\centering
\caption{Signal efficiencies (\%), and expected signal and background yields
for $\mes>5.27 \gevcc$, for low and high $q^2$ regions.}
\begin{tabular}{|l|rr|rr|rr|} \hline\hline
     & \multicolumn{2}{c|}{Signal Eff.} & \multicolumn{2}{c|}{Signal Yield} & \multicolumn{2}{c|}{Bkgd. Yield} \\
Mode                & low  & high            & low  & high                     & low  & high \\ \hline
$\modesixshort$     & 1.6 & 3.1              & 1.0  & 1.8                      & 0.7 & 3.8   \\
$\modesevenshort$   & 3.6 & 5.5              & 3.0  & 4.5                      & 0.3 & 1.4   \\
$\modeeightshort$   & 4.5 & 8.1              & 5.5  & 9.6                      & 0.0 & 3.1   \\
$\modetenshort$     & 4.6 & 5.3              & 2.8  & 3.1                      & 1.7 & 2.4   \\
$\modeelevenshort$  & 7.0 & 5.4              & 5.9  & 4.4                      & 0.3 & 1.4   \\
$\modetwelveshort$  & 8.6 & 10.3             & 10.5 & 12.2                     & 1.7 & 2.4   \\
\hline
Total Yield         &     &                  & 28.6 & 35.8                     & 4.8 & 14.5  \\
\hline\hline
\end{tabular}
\label{tab:effs}
\end{table}

For each $q^2$ region, we combine events from all six final states
and perform three successive unbinned maximum likelihood fits.
Because of the relatively small number of signal
candidates in each $q^2$ region, a simultaneous
fit over $\mes$, $\ctk$ and $\ctl$ is unlikely to converge
and a sequential fitting procedure is required.
We initially fit the $\mes$ distribution using events
with $\mes > 5.2 \gevcc$ to obtain the signal and
background yields, $N_S$ and $N_B$ respectively.
We use an ARGUS shape~\cite{ArgusShape} with a free shape parameter
to describe the combinatorial background in this fit.
For the signal, we use a Gaussian shape with a mean
$\mes = 5.2791 \pm 0.0001\gevcc$ and $\sigma = 2.60 \pm 0.03\mevcc$,
which are determined from a fit to the vetoed charmonium samples.
In this and subsequent fits we account for a small contribution
from misidentified hadrons by subtracting the
$\Kstar h^{\pm}\mu^{\mp}$ events, weighted by the probability for the
$h^{\pm}$ to be misidentified as a muon.
We also account in all fits for charmonium events that escape the veto,
and for mis-reconstructed signal events.
We estimate contributions from non-resonant $K\pi$ decays
by fitting events outside the $\Kstar$ mass window in the range $0.7-1.1 \gevcc$.
We find no signal-like events that are not accounted for by the tails of the
resonant mass distribution, and thus do not expect any significant contribution
from non-resonant events within the mass window.

The second fit is to the cosine of the
helicity angle of the $\Kstar$ decay, $\ctk$,
for events with $\mes > 5.27 \gevcc$.
In this fit, the only free parameter is $\fl$,
with the normalizations for signal and combinatorial
background events taken from the initial $\mes$ fit.
The background normalization is obtained by integrating,
for $\mes > 5.27 \gevcc$, the ARGUS shape resulting from the $\mes$ fit.
We model the $\ctk$ shape of the combinatorial background
using \epem\ and \mumu\ events, as well as lepton-flavor violating \emu
and \mue\ events, in the $5.20 < \mes < 5.27 \gevcc$ sideband.
The signal distribution given in equation (1) is folded
with the detector acceptance as a function of
$\ctk$, which is obtained from simulated signal events.

The final fit is to the cosine of the lepton helicity angle, $\ctl$,
for events with $\mes > 5.27 \gevcc$.
The only free parameter in this fit is $\afb$,
with the signal distribution given in equation~(2)
folded with the detector acceptance as a function
of $\ctl$.
In this fit, the value
of $\fl$ is fixed from the result of the second fit, and normalizations
for signal and combinatorial background events are identical
to those used in the second fit.
We constrain the $\ctl$ shape of the combinatorial background
using the same sideband samples as for the $\ctk$ fit.
The correlated leptons from $B\to D^{(*)}\ell\nu$,
$D\to K^{(*)}\ell\nu$ give rise to an $\mes$-dependent peak
in the combinatorial background at $\ctl>0.7$, and
we consider this correlation in our study of systematic errors.
No such correlation is observed for $\ctk$.

\begin{table}[tb!]
\centering
\caption{Results for the $B\to J/\psi K^*$ control samples. $\Delta$BF are the differences
between the measured branching fractions and the world average value~\cite{JPsi_BF}.
The previously measured $\fl = 0.56 \pm 0.01$~\cite{JPsi_FL},
and the expected $\afb=0$.}
\begin{tabular}{cccc} \hline\hline
Mode & $\Delta$BF ($10^{-3}$) & $\fl$ & $\afb$ \\ \hline
$K^+\pi^0\mu^+\mu^-$ & $+0.09\pm 0.12$ & $0.54\pm 0.03$ & $-0.04\pm 0.05$\\
$K_S^0\pi^+\mu^+\mu^-$ & $+0.02\pm 0.11$ & $0.55\pm 0.02$ & $+0.00\pm 0.05$\\
$K^+\pi^-\mu^+\mu^-$ & $-0.03\pm 0.07$ & $0.56\pm 0.02$ & $-0.02\pm 0.02$ \\
$K^+\pi^0e^+e^-$ & $+0.16\pm 0.10$ & $0.54\pm 0.03$ & $+0.02\pm 0.03$ \\
$K_S^0\pi^+e^+e^-$ & $+0.07\pm 0.10$ & $0.55\pm 0.02$ & $-0.02\pm 0.04$ \\
$K^+\pi^-e^+e^-$ & $+0.02\pm 0.07$ & $0.56\pm 0.02$ & $+0.01\pm 0.02$ \\
\hline\hline
\end{tabular}
\label{tab:jpsibfs}
\end{table}

We test our fits using the large
sample of vetoed charmonium events.
The branching fractions (BF) and $K^*$ polarization for
$B\to J/\psi K^*$ are well known~\cite{JPsi_BF, JPsi_FL},
and $\afb$ is expected to be zero. The results of the fits to
the six final states are all consistent with expected values
(see Table~\ref{tab:jpsibfs}).
We further test our methodology by performing the
$\mes$ and $\ctl$ fits on a sample of $B^+\to K^+\ellell$ decays.
The results are given in Table~\ref{tab:kstll} and
are consistent with negligible forward-backward asymmetry,
as expected in the SM and most new physics models~\cite{Bobeth:2007dw}.

\begin{table}[tb!]
\centering
\caption{Results for the fits to the $K\ellell$ and $K^*\ellell$ samples.
$N_S$ is the number of signal events in the $\mes$ fit.
The quoted errors are statistical only.}
\begin{tabular}{lcccc} \hline\hline
Decay & $q^2$  & $N_S$ &  $\fl$ & $\afb$ \\ \hline \vspace{-.1in}\\ \vspace{.04in}
$K\ellell$ & low & $26.0\pm 5.7$ & & $+0.04^{+0.16}_{-0.24}$\\ \vspace{.04in}
& high & $26.5\pm 6.7$ &  & $+0.20^{+0.14}_{-0.22}$ \\ \hline\vspace{-.1in}\\ \vspace{.04in}
$K^*\ellell$ &  low  & 27.2 $\pm$ 6.3 &  $0.35\pm 0.16$ & $+0.24^{+0.18}_{-0.23}$ \\\vspace{.04in}
& high  & 36.6 $\pm$ 9.6 & $0.71^{+0.20}_{-0.22}$ & $+0.76^{+0.52}_{-0.32}$ \\
\hline\hline
\end{tabular}
\label{tab:kstll}
\end{table}

We validate the fit model by performing ensembles of fits
to datasets with events drawn from simulated signal and
background event samples. The input SM values of $\fl$ and $\afb$
are reproduced with the expected statistical errors.
A few percent of the fits do not converge due to small signal yields.
We have also performed fits using signal events
generated with widely varying values of $C_7^{\textrm{\tiny{eff}}}$,
$C_9^{\textrm{\tiny{eff}}}$ and $C_{10}^{\textrm{\tiny{eff}}}$
covering the physically allowed regions of $F_L$ and $\afb$,
and find minimal bias in our fits.

\begin{table}[tb!]
\centering
\caption{Systematic errors on the measurements of $F_L$ and $\afb$ in
the $K^*\ellell$ samples.}
\begin{tabular}{lcccc} \hline\hline
Source & \multicolumn{2}{c}{$\fl$} & \multicolumn{2}{c}{$\afb$} \\
of Error & low $q^2$ & high $q^2$ & low $q^2$ & high $q^2$\\ \hline
$\mes$ fit yields     & 0.001 & 0.016 & 0.003 & 0.002 \\
$\fl$ fit error       &       &       & 0.025 & 0.022 \\
Background shape      & 0.011 & 0.008 & 0.017 & 0.021 \\
Signal model          & 0.036 & 0.034 & 0.030 & 0.038 \\
Fit bias              & 0.012 & 0.020 & 0.023 & 0.052 \\
Mis-reconstructed signal & 0.010 & 0.010 & 0.020 & 0.020 \\ \hline
Total                 & 0.041 & 0.044 & 0.052 & 0.074 \\
\hline\hline
\end{tabular}
\label{tab:syst}
\end{table}

The systematic errors on the fitted values of $\fl$ and $\afb$
are summarized in Table~\ref{tab:syst}.
The uncertainties in the fitted signal yields $N_S$,
due to variations in the ARGUS shape in the $\mes$ fits,
are propagated into the angular fits.
The errors on the fitted $\fl$ values are propagated into the $\afb$ fits.
We vary the combinatorial background shapes by dividing the sideband sample
into two disjoint regions in $\mes$. We vary the signal model using
simulated events generated with different form factors~\cite{AFB_SM, BallZwicky},
and with a range of values of $C_7^{\textrm{\tiny{eff}}}$,
$C_9^{\textrm{\tiny{eff}}}$ and $C_{10}^{\textrm{\tiny{eff}}}$,
to determine an average fit bias.
Finally, the modeling of mis-reconstructed signal
events is constrained by the fits to the charmonium samples (Table~\ref{tab:jpsibfs}), where
it is the largest systematic uncertainty.

\begin{figure}[tb!]
\begin{center}
\includegraphics[width=1.0\linewidth]{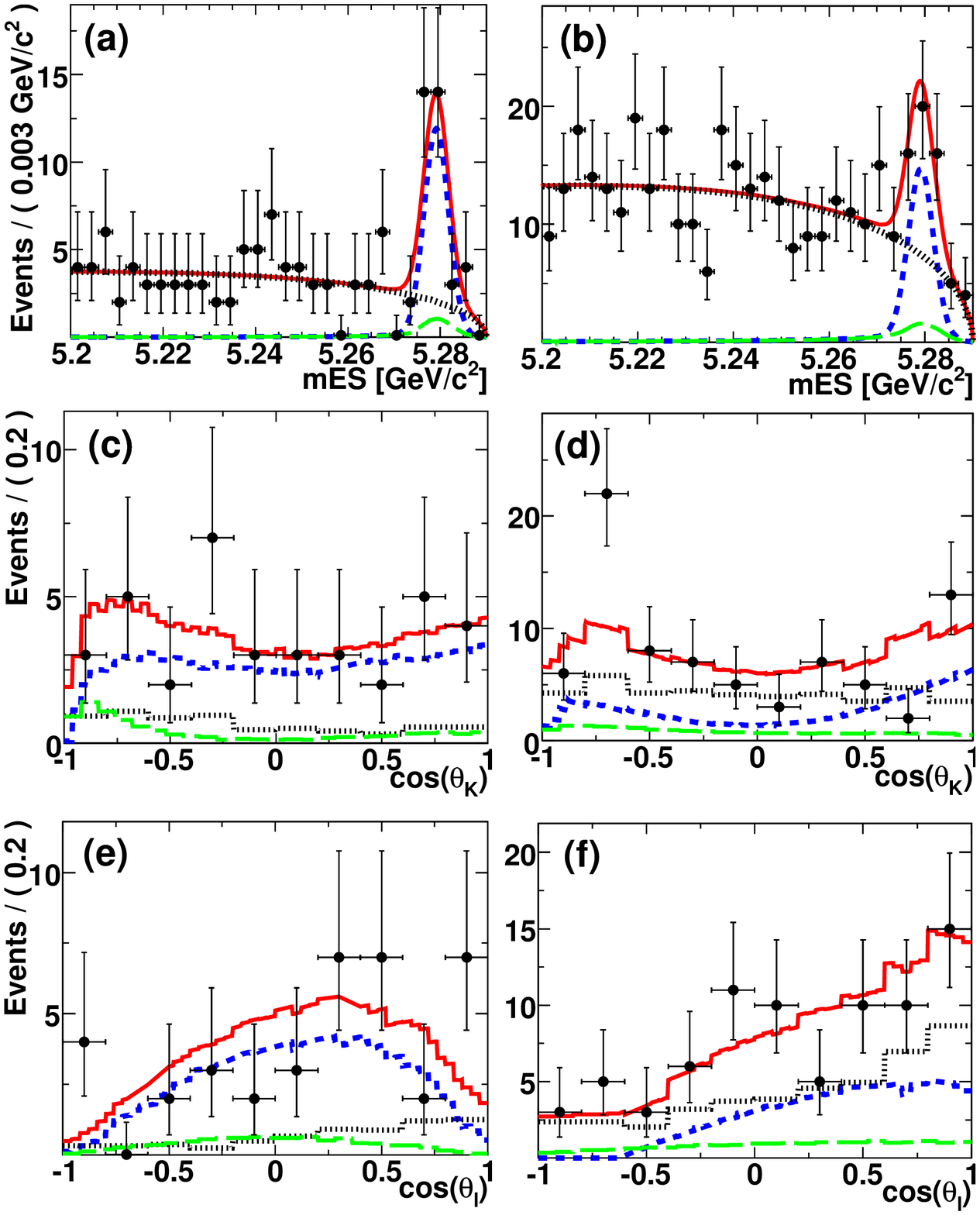}
\caption{$K^*\ellell$ fits:
(a) low $q^2$ \mes,
(b) high $q^2$ \mes,
(c) low $q^2$ \ctk,
(d) high $q^2$ \ctk,
(e) low $q^2$ \ctl,
(f) high $q^2$ \ctl;
with combinatorial (dots) and
peaking (long dash) background,
signal (short dash) and total (solid)
fit distributions superimposed
on the data points.}
\label{fig:fitplots}
\end{center}
\end{figure}

The final fits to the $K^*\ellell$ samples are shown in
Fig.~\ref{fig:fitplots}. The results for $\fl$ and $\afb$
are given in Table~\ref{tab:kstll}
and are shown in Fig.~\ref{fig:newafbfl}.
In the low $q^2$ region, where we expect $\afb \sim -0.03$
and $\fl \sim 0.63$ from the SM,
we measure $\afb=0.24^{+0.18}_{-0.23} \pm 0.05$
and $\fl=0.35 \pm 0.16 \pm 0.04$, where the first error
is statistical and the second is systematic.
In the high $q^2$ region, the SM expectation is
$\afb \sim 0.38$ and $\fl \sim 0.40$,
and we measure $\afb=0.76^{+0.52}_{-0.32} \pm 0.07$
and $\fl=0.71^{+0.20}_{-0.22} \pm 0.04$,
with a signal yield of $36.6 \pm 9.6$ events.
Theoretical uncertainties on the expected SM $\fl$ and $\afb$ values
are generally difficult to characterize in the high $q^2$ region, and
although under better control for $1<q^2<6\gevcccc$, the
extension of our low $q^2$ region below $1\gevcccc$ makes estimates
of uncertainties there difficult also.
The quoted values are obtained using our
implementation of the physics models described
in~\cite{Ali01, BallZwicky}, corresponding to the
SM curves in~Fig.~\ref{fig:newafbfl}.

The expected SM value of $C_{10}^{\textrm{\tiny{eff}}}$ at next-to-next-to-leading
logarithmic (NNLL) order is
$C_{10}^{\textrm{\tiny{eff}}} = -4.43$~\cite{Beneke:2004dp}.
A more recent NNLL calculation which evaluates contributions from the full set
of seven form factors gives $C_{10}^{\textrm{\tiny{eff}}} = -4.13$~\cite{Altmannshofer:2008dz}.
The magnitude of possible contributions from new physics to $C_{10}$ can
be constrained if $\afb>0$ at high $q^2$.
By combining such a constraint on $\afb$ with inclusive $b\rightarrow s\ellell$ branching fraction results,
an upper bound of $|C_{10}^{NP}| \lesssim 7$ can be obtained,
improving on an upper bound derived solely from branching
fraction results of $|C_{10}^{NP}| \lesssim 10$~\cite{Bobeth:2008ij}.
Our results are consistent with measurements by
Belle~\cite{Belle_afb}, and replace the earlier
\babar\ results in which only a lower limit on $\afb$ was set in
the low $q^2$ region~\cite{BaBar_Kll}.

We are grateful for the excellent luminosity and machine conditions
provided by our \pep2\ colleagues,
and for the substantial dedicated effort from
the computing organizations that support \babar.
The collaborating institutions wish to thank
SLAC for its support and kind hospitality.
This work is supported by
DOE
and NSF (USA),
NSERC (Canada),
CEA and
CNRS-IN2P3
(France),
BMBF and DFG
(Germany),
INFN (Italy),
FOM (The Netherlands),
NFR (Norway),
MES (Russia),
MEC (Spain), and
STFC (United Kingdom).
Individuals have received support from the
Marie Curie EIF (European Union) and
the A.~P.~Sloan Foundation.

\end{document}